\documentclass[aps,pre]{revtex4}
\usepackage{graphicx}% Include figure files
\usepackage{color}

\def\be{\begin{equation}}
\def\ee{\end{equation}}     
\def\bfi{\begin{figure}}
\def\efi{\end{figure}}
\def\bea{\begin{eqnarray}}
\def\eea{\end{eqnarray}}

\begin{document}

\title{Phase ordering in disordered and inhomogeneous systems}

\author{Federico Corberi}
\affiliation {Dipartimento di Fisica ``E.~R. Caianiello'', and INFN, Gruppo Collegato di Salerno, and CNISM, Unit\`a di Salerno,Universit\`a  di Salerno, 
via Giovanni Paolo II 132, 84084 Fisciano (SA), Italy.}

\author{Eugenio Lippiello}
\affiliation{Department of Mathematics and Physics, Second University of Naples, Viale Lincoln 5, 81100 Caserta, Italy}

\author{Raffaella Burioni}
\affiliation {Dipartimento di Fisica e Scienza della Terra, and INFN, Gruppo 
Collegato di Parma, Universit\`a di Parma, Parco Area delle Scienze 7/A, I-423100 
Parma, Italy.}

\author{Alessandro Vezzani}
\affiliation{Centro S3, CNR-Istituto di Nanoscienze, Via Campi 213A, 41125 Modena Italy, and Dipartimento di Fisica Scienza della Terra, Universit\`a di Parma, Parco Area delle Scienze 7/A, I-43100 Parma, Italy.}

\author{Marco Zannetti}
\affiliation {Dipartimento di Fisica ``E.~R. Caianiello'', and INFN, Gruppo Collegato di Salerno, and CNISM, Unit\`a di Salerno,Universit\`a  di Salerno, via Giovanni Paolo II 132, 84084 Fisciano (SA), Italy.}

\begin{abstract}

We study numerically the coarsening dynamics of the Ising model 
on a regular lattice with random bonds and on deterministic fractal substrates.
We propose a
unifying interpretation of the phase-ordering processes 
based on two classes of dynamical behaviors 
characterized by different growth-laws 
of the ordered domains size - logarithmic or
power-law respectively. It is conjectured that the interplay between
these dynamical classes is regulated by
the same topological feature which governs the presence or the absence of a
finite-temperature phase-transition.

\end{abstract}

\maketitle

\section{Introduction}

When a ferromagnetic system is quenched to below the critical temperature
a slow non-equilibrium phase-ordering process takes
place with domains of the ordered phases increasing their size $L(t)$ in 
time \cite{bray,2dinoirf}. Typical examples are ferromagnets,
binary liquids or alloys. When time is large, due to the presence of the
dominant length scale $L(t)$ dynamical
scaling sets in~\cite{zan,BCKM,ioeleti}. 
The main features of this structure, which is quite well understood in 
non-disordered homogeneous systems, is expected to be valid also 
in the presence of quenched disordered or when homogeneity is spoiled. 
Recently, this has promoted a considerable effort to understand coarsening
phenomena in disordered and in non-homogeneous 
systems~\cite{diluted,sab08,puripowerlaw,
super2,hp06,
lastparma,super3,decandia,EPL,CLMPZ,variousnoi2,various2,super1,10noirf,pp,HH,parma}.
Particular attention was devoted to the asymptotic growth law
since, although a logarithmic behavior
was generally expected, also a power-law sometimes was reported \cite{puripowerlaw,super2,hp06}. 
 
In a recent paper \cite{diluted} some of us have studied a system - the site diluted Ising model 
(SDIM) - 
where, the interplay between logarithmic and power-law
growth can be fully understood and the occurrence of the two types of behavior
can be tuned by means of the amount of dilution. In the site diluted model 
Ising spins are located on a substrate which is obtained from a 
regular  lattice by removing randomly a
fraction $D$ of sites. 
In the 
{\it pure} case with $D = 0$, the usual temperature-independent
power-law $L(t)\sim t^{1/z}$ is obeyed, where $z=2$ for a dynamics
without conservation of the order parameter, as it will be considered here.
Increasing $D$, a region with an 
asymptotic logarithmic behavior of $L(t)$ is entered. However the situation
changes when the critical value $D=D_c$ is reached, such that the fraction $P_c=1-D_c$ 
of spins is at the percolation threshold. In this case a temperature-dependent 
power-law
$L(t)\sim t^{1/\zeta (T)}$ is observed. This is interpreted as due to the different
topology of the substrate, for $D<D_c$ and for $D=D_c$, since in the former
case it is compact on large scales, while in the latter it is a percolation fractal.
Notice that the critical temperature $T_c$ of the model, which is finite
for $D<D_c$, vanishes at the percolation threshold $D=D_c$. 

The role of the substrate topology was also explored, in a different context,
in \cite{lastparma} where the coarsening of the Ising model
defined on deterministic fractal networks was considered. 
There it was shown that, again,
two types of growth, logarithmic versus temperature dependent power-law,
could be observed depending on the substrate considered. In particular,
it was argued that logarithmic behavior is found on networks with a finite
Ising critical temperature, while temperature-dependent power laws are observed on
structures with $T_c=0$. 
 
The above findings suggest that 
the substrate topology, which is responsible for the existence of the equilibrium phase-transition,
might also be important in determining the non-equilibrium growth-law, leading to the conjecture
that temperature-dependent power-laws are to be expected on
inhomogeneous or disordered systems with $T_c=0$ -
like the SDIM at the percolation threshold - while 
a logarithmic behavior occurs when $T_c>0$, as in the same model with $D<D_c$.

In the above mentioned cases, 
the Ising model is constructed on a substrate with topological properties
obtained by diluting the lattice either 
randomly or according to a deterministic rule.  
Other thoroughly studied systems such as the Ising model with random
bonds or random fields are defined on homogeneous lattices. In these cases 
disorder pins the interfaces whose evolution becomes
site-dependent and the growth-law is observed to be
much slower than in the corresponding pure systems
\cite{decandia,EPL,CLMPZ,variousnoi2}.

In this Article, by studying an asymmetric version of the 
random-bond Ising model, we argue that this is the case. 
We explicitly exhibit the parameters controlling the speed of growth and set them 
as to produce either a faster or a slower asymptotic growth law which are well compatible
with an algebraic or a logarithmic behavior.  
These parameters have a clear geometrical interpretation in terms
of topological properties of the bond network and, accordingly, one finds 
logarithmic or temperature-dependent power-law growths. 
Our results hint to promote the relation 
between topology and growth-law to a general feature of inhomogeneous 
and/or disordered phase-ordering systems. 

The paper is organized as follows. The random bond model is defined in Sec. \ref{GRBIM}.
Numerical simulations are discussed in Sec. \ref{SIMUL}.
In order to better understand the role
of topology, a pair of systems with Ising spins 
on deterministic fractal substrates are introduced in  Sec. \ref{GAS}. The
study of these models allows to gain useful qualitative insights into
the more complicated random bond system.
Finally, some open issues are briefly discussed in the concluding Sec. \ref{CONCL}.

\section{The asymmetric Random Bond Ising Model}
\label{GRBIM}

We consider an Ising model,
hereafter referred to as the asymmetric random-bond Ising model (ARBIM)
which is described by the following Hamiltonian
\be
H=-\sum _{<ij>}J_{ij}\sigma _i \sigma _j.
\label{ham}
\ee
Here $\sigma _i=\pm 1$ are spins on a 2-d square lattice, $<ij>$ are nearest
neighbors and $J_{ij}=J_0+\xi _{ij}$ are positive coupling constants. 
The $\xi _{ij}$ are uncorrelated random variables that can take two values
$\pm \epsilon$, with $\epsilon \le J_0$
to ensure the positivity of $J_{ij}$. 
The probability of occurrence $P(\xi _{ij})$ of the two possible values 
$\xi _{ij}=\pm \epsilon$ is assumed to be in general asymmetric with
$P(\xi _{ij}=-\epsilon)=d$.
In order to be concrete, 
let us make the example of a case which will be important in the 
following, with $\epsilon =J_0$. 
This situation corresponds to a model with random coupling constants 
bi-modally distributed on the two possible values
$J_{ij}=0$ and $J_{ij}=2J_0$ occurring with probability $d$ and $1-d$, respectively. 
Therefore, this case corresponds to the Ising model
with $J_{ij}=2J_0$ and bond dilution, namely with a fraction $d$ of the 
bonds removed.

\subsection{Space of parameters}

\begin{figure}[h]
  \vspace{1cm}
    \centering
    %\vbox to 8.5 cm {                                                          
   \rotatebox{0}{\resizebox{.45\textwidth}{!}{\includegraphics{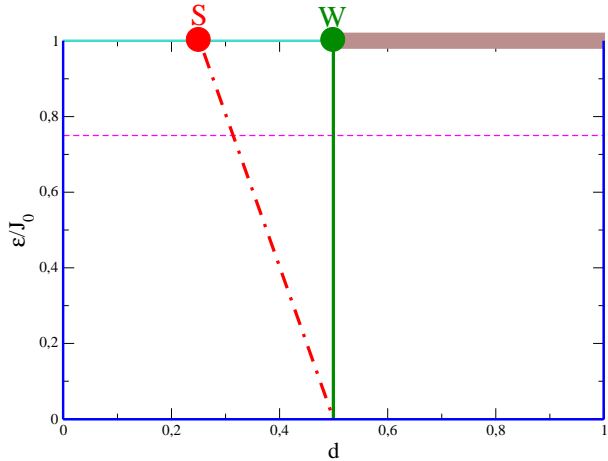}}}
   \caption{(Color online). Parameter space of of the model. The blue lines
   on the left, right and bottom of the figure correspond to a pure system.
   Red and green heavy dots are the strong-disorder (S) and percolative (W)
   fixed points with $\epsilon =J_0$. Dot-dashed red and continuous green 
   heavy lines are the
   line of fixed points originating from S and W upon changing $\epsilon$
   (see Secs. \ref{epslt1m2},\ref{firstsum}).}
\label{phas_diag}
\end{figure}

In the limit of low temperatures that we will always consider in this paper,
the model depends only on $d$ and $\epsilon$.
In the parameter space, sketched in Fig.~\ref{phas_diag}, 
only the region with $0\le d\le 1$, 
$0\le \epsilon /J_0 \le 1$
is allowed since $d$ is a probability and because we limit our analysis to non-negative 
coupling constants in order to avoid the different problem where frustration
occurs (notice that we 
use $\epsilon /J_0$ on the diagram axis instead of $\epsilon $).

In Fig. \ref{phas_diag} the axes with $d=0$, $d=1$, and $\epsilon =0$ 
correspond to the pure Ising model (i.e. there is no disorder in the couplings).
This is pictorially represented by drawing them in blue. 

The point $d=1/2, \epsilon /J_0 =1$, hereafter referred to as $W$, will be 
particularly relevant in the following since, recalling the discussion above, it 
can be regarded as a system with uniform couplings $J_{ij}\equiv 2J_0$ 
but where a fraction 
$d=d_c=1/2$
of the bonds has been randomly removed. Since $p_c=1-d_c=1/2$ is 
the bond percolation threshold, the system is at the percolation point of the
substrate. 
Let us recall here that the Ising model defined on this percolative
network is characterized by $T_c=0$. This is due to the presence of the
so called {\it cutting bonds} \cite{Stauffer}, namely isolated bonds whose removal would
cause the disconnection of arbitrarily large parts of the structure. 
In this sense the network is 
{\it weak} and we denote this point with the letter $W$.
Notice that the region with 
$\epsilon /J_0 =1,d>d_c$, represented in brown
in Fig. \ref{phas_diag}, is not interesting since on that 
sector the substrate is disconnected and asymptotic phase-ordering cannot occur.

\subsection{Time Evolution}
\label{timev}

We implement non-conserved dynamics\cite{bray,2dinoirf} by evolving the spins 
with a single-spin-flip transition rates of the Glauber form
\be
w(\sigma _i \to -\sigma _i)=\frac{1}{2} \left [1-\sigma _i \tanh (H_i^W/T)\right ] .
\label{transr}
\ee
Here, $H_i^W$ is the local Weiss field obtained by the sum
\be
H_i^W = \sum _{j \in L_i} J_{ij} \sigma_j
\ee
over the set of nearest-neighbors $L_i$ of $i$. 

Before presenting numerical simulations, in the next section we briefly overview
the behavior of the closely related site diluted model.

%\subsubsection{The symmetric case $d=1/2$ with uniform distribution of coupling constants.}
%\label{unif}

%The case in which the coupling constants are symmetrically 
%and uniformly distributed (instead of having a bimodal distribution
%as in in the present ARBIM) in 
%the interval $[J_0-\epsilon,J_0+\epsilon]$ 
%has been thoroughly studied in the past \cite{puripowerlaw,super2,CLMPZ}.
%In this case one observes a long-lasting regime
%with a power-law growth. 
%Recently \cite{CLMPZ} it was shown that this stage is only preasymptotic,
%since the analysis of very clean data for the effective exponent
%clearly shows an incipient crossover to a logarithmic law at the longest
%times accessible in the simulations.

\subsubsection{The segment $\epsilon =J_0$ and the relation with the site
diluted model} \label{e1m2}

It is useful to stress the close relation between the segment $\epsilon /J_0=1$,
$0\le d \le d_c$ of the present model and the
SDIM studied in \cite{diluted}. The latter has a fixed coupling constant 
but a  fraction of spins is randomly removed with a probability $D$
(we use capital $D$ to distinguish it from the ARBIM disorder parameter $d$). 
ARBIM with $\epsilon /J_0 =1$ differs from SDIM because of bond dilution in place of site
dilution.
Since we expect similar
properties for the two systems it is useful to overview first the behavior
of the SDIM. In this model the dilution parameter 
can be varied between 
$D=0$, corresponding to the pure Ising model, up to $D_c=1-P_c$, where
$P_c\simeq 0.5928$ is the site percolation threshold. For $D>D_c$ the substrate
is disconnected and asymptotic coarsening cannot occur, not differently
from the ARBIM for $d$ above $d_c$. 

In Ref.\cite{diluted} it was shown that
the SDIM kinetic properties display a scaling structure
which is most effectively described in terms of three competing fixed points
on the $D$ axis, making free use of renormalization group terminology. The first is the trivial
one of the pure system, located at $D=0$, and is associated to the usual
power-law growth $L(t)\sim t^{1/2}$. The second one is the {\it percolative}
fixed point located at $D_c$ and is characterized by a temperature-dependent
power-law growth $L(t)\sim t^{1/\zeta (T)}$. These fixed points are repulsive,
in the sense that as soon as $D > 0$, or $D < D_c$ the asymptotic dynamics
is governed by a different fixed point, located at $D=D^*\simeq 0.225$, 
to whom a logarithmic increase of $L(t)$ is associated. Due to this fixed point 
structure, if the system is prepared with an intermediate dilution 
between $D=0$ and $D^*$ (or equivalently between $D^*$ and $D_c$)
a crossover is observed at a certain time $t_{cross}(T,D)$ from an initial transient regime 
governed by the nearest unstable fixed point ($D=0$ or $D_c$ respectively),
with a power-law increase of the domains size, to a late regime controlled
by the attractive point at $D^*$, characterized by a logarithmic $L(t)$.
As a result, the slowest possible growth, namely the one where $t_{cross}(T,D)$ is
smallest, is obtained at $D=D^*$ and 
comparing $L(t)$ for different values of $D$ one finds a non-monotonous 
behavior: The growth
slows down in going from $D=0$ to $D=D^*$ and then speeds up again
when $D$ is further increased from $D=D^*$ up to $D=D_c$. 
This can be used as a practical way to identify $D^*$.

Since the only difference between the ARBIM with $\epsilon/J_0=1$
and the SDIM
is bond dilution replacing site dilution, we expect to observe the same
pattern with an attractive
fixed point associated to logarithmic growth and located somewhere in
between $d=0$ and $d=d_c$. This is represented in Fig. \ref{phas_diag}
by the heavy red dot marked with the letter $S$, corresponding to the
dilution $d=d^*$. For this value of $d$ the growth-law is expected to be
the slowest possible one, in the sense discussed above.
Evidence for the existence of this fixed point  will be discussed in Sec. 
\ref{SIMUL}. 

\section{Numerical simulations} \label{SIMUL}

We have performed a series of simulations of the ARBIM
by considering a cooling procedure 
where the system is prepared initially in the infinite-temperature disordered state
and, at the time $t=0$, it is suddenly quenched to a finite temperature $T$.
If not specified differently we use $T=\epsilon$.
This temperature is chosen as a compromise between the aim of
studying the low temperature sector, where dynamical properties can be
better interpreted, and the severe limits posed by the slowing down of the dynamics for $T\to 0$.

In order to speed up simulations we have used a modification of the dynamics 
introduced in Sec. \ref{timev} where flipping spins in the 
bulk of domains, namely those aligned with all the nearest neighbors, is 
prevented. This modified dynamics does not alter the behavior of the quantities we 
are interested in, as it has been tested in a large number of 
cases \cite{nobulk,Ontheconnec,diluted}. We 
have checked that this is true also in the present study.
 
In our simulations we have used sufficiently large system sizes in order not to have
any detectable finite-size effect in the range of time accessed in the runs.
Specifically, we have considered a two-dimensional square lattice 
of $N={\cal L}^2$ sites with ${\cal L}$ in the range $[10^3 - 2\cdot 10^3]$
for quenches in different sectors of the parameter space of Fig. \ref{phas_diag}. 
For every choice of the parameters, we have performed a certain number 
(in the range 10-100) of independent 
runs with different initial conditions and thermal histories in order to populate the 
non-equilibrium ensemble needed to extract the average quantities that will be introduced below. 

The observable of interest in this paper is the typical domains size $L(t)$.
In homogeneous systems this quantity is trivially related by 
\be
L(t)=e ^{-1}(t)
\label{lt}
\ee
to the excess energy density 
\be
e(t)=\frac{1}{V}[<H(t)>-<H>_f],
\ee
where $<H>_f$ is the average energy of the equilibrium state at the final temperature of the quench.
Eq. (\ref{lt}) simply states that the excess energy is stored on the interfaces whose density scales as
the inverse of the size of the growing domains \cite{bray}.

In the model considered here the substrate can have disconnected parts. This happens, for instance, for 
$\epsilon =J_0$ and sufficiently large values of $d$.
In this case phase 
ordering occurs independently and with different characteristics on the various parts of the system
and, correspondingly, different definitions of the growing length can be given. 
Let us consider the case $\epsilon =J_0$, $d=1/2$, where the substrate if formed by an infinite spanning 
cluster and many finite-size parts.
Since we are interested in the aging 
phenomenon related to the existence of a divergent length, one would define 
$L(t,d)$ as the characteristic length of the ordered regions which are effectively 
growing. 
Keeping in mind the example with $\epsilon =J_0$ one can argue that the quantity (\ref{lt}) 
introduced before is suited to the task. 
Indeed, at any given time there will be a number of sufficiently small disconnected parts 
of the substrate which are already ordered. These pieces of the system do not contribute 
to the computation of $L(t,d)$ in Eq. (\ref{lt}), because a finite cluster is by definition 
surrounded by bonds with $J_{ij}=J_0=\epsilon=0$ 
and hence there is no excess energy associated with it when its inner spins are aligned. 
For this reason in this paper we use the determination (\ref{lt}) of
$L(t)$.  
A discussion of the relation between Eq. (\ref{lt}) and other possible definitions of L(t) in non-homogeneous systems can be found in Ref. \cite{diluted}.

In the following we will describe the behavior of the ARBIM in different
regions of the parameter space.

\subsection{$\epsilon=J_0$} \label{eps1m2}

We start by investigating the line $\epsilon /J_0=1$, by varying $d$.
This corresponds to scan the segment represented in turquoise in Fig. \ref{phas_diag}.  
As discussed in Sec. \ref{e1m2} in this region of the parameter space we expect 
to see a behavior analogous to that observed in the SDIM. Specifically,
as the parameter $d$ is varied, one should
observe power-laws for $L(t)$ when $d$ is set to the limiting values $d=0$ and 
$d=d_c=1/2$, a logarithmic behavior for a certain value $0<d^*<d_c$, and
a crossover from power-law to logarithm for intermediate values of $d$. 
Hence, comparing $L(t)$ for different values of $d$ 
a non-monotonous behavior is expected, with a faster growth at $d=0$
progressively slowing down upon approaching $d=d^*$, and then 
speeding up again in going from $d=d^*$ to $d=d_c$.

\begin{figure}[h]
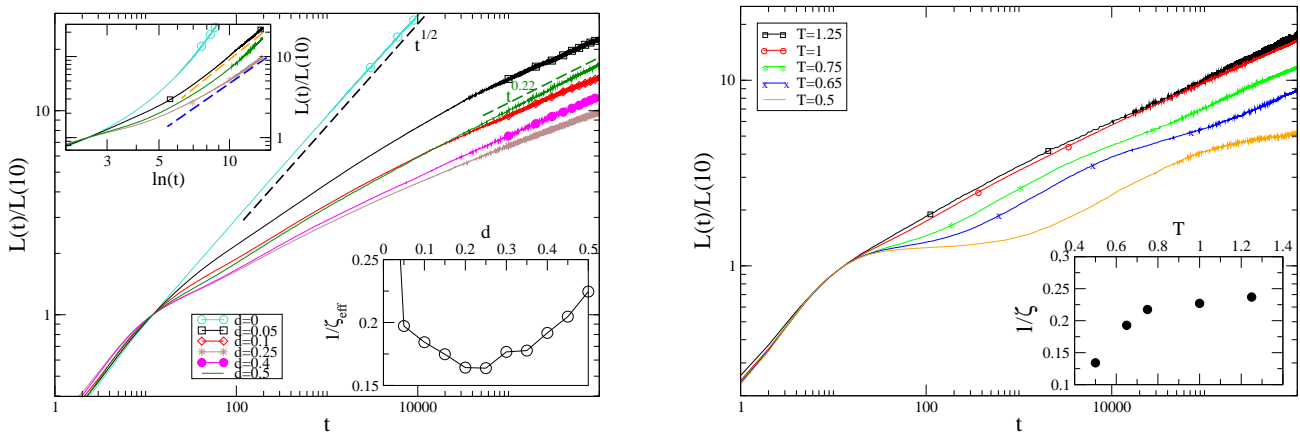

  \vspace{1cm}
    \centering
    %\vbox to 8.5 cm {                                                          
\rotatebox{0}{\resizebox{.44\textwidth}{!}{\includegraphics{m2_eps1.eps}}}
\hspace{1cm}
\rotatebox{0}{\resizebox{.45\textwidth}{!}{\includegraphics{m2_eps1_varieT.eps}}}
\caption{(Color online). Panel (a) (left) :$L(t)$ is plotted against $t$ in a double-log plot 
for a quench of the ARBIM to $T=\epsilon$, with $\epsilon =J_0$ 
and different values of $d$ specified in the key. The black-dashed line is the
pure-case power-law $L(t)\propto t^{1/2}$ and the green-dashed one
is the fit $L(t)\sim t^{0.22}$ at $d=d_c$. 
In the upper inset some of the curves of the main picture are plotted against
$\ln t$ on a double-log plot. The orange-dashed and blue-dashed lines are 
the fits $y=0.03 \cdot [\ln t]^{2.5}$ and $y=0.04 \cdot [\ln t]^{2.1}$, respectively.
In the inset the effective
exponent $1/\zeta _{eff}$ is plotted against $d$.
Panel (b) (right): $L(t)$ is plotted against $t$ in a double-log plot
for a quench of the ARBIM to different final temperatures (specified in the key,
curves lower upon decreasing temperature), 
with $\epsilon =J_0$ and $d=0.5$. In the inset the exponent $1/\zeta (\epsilon=J_0,T,d=d_c)$,
obtained from a fit of the curves from $t=10^2$ onwards, is plotted against $T$.}
%Fits: Black: $0.029408*x^2.5269; Red: 0.029239*x^2.3642; Brown: 0.03965*x^2.099; 
%Magenta: 0.018442*x^2.4575; Green: 0.0087453*x^2.8778$
\label{m2_eps1}
\end{figure}

As shown in the left panel of Fig. \ref{m2_eps1} this is precisely what one observes in the ARBIM.
Notice that here and in the following, in order to better compare the asymptotic
behavior of $L(t)$ as the parameters are changed we plot $L(t)/L(10)$
to make the curves cross at the time $t_{early}\simeq 10$ when the early
regime is over.  
In the case $d=0$ one recovers the behavior $L(t)\sim t^{1/2}$ of the pure case.
As $d$ is increased up to $d^*\simeq 0.25$, $L(t)$ gets slower and slower but
it grows faster again above $d^*$. Notice that at $d=d_c$ one has a behavior
compatible with a power-law $L(t)\sim t^{1/\zeta(\epsilon,T,d)}$, with 
$1/\zeta(\epsilon=J_0,T=\epsilon,d_c)\simeq 0.22$ in this case.
The same power-law behavior is observed also by quenching to different temperatures,
as it is shown in the right panel of Fig. \ref{m2_eps1}. Here it is shown that 
the growth is slower for lower temperatures, signaling that $\zeta$ depends on $T$.
Notice also that, for low temperatures, the growth of $L(t)$ is decorated by periodic
oscillations which can be interpreted as due to the recurrent trapping of the interfaces
on the pinning centers.
The value of $\zeta $, extracted by fitting the data
with a power-law for $t\ge 10^2$ is shown in the inset. Here it is observed that 
$1/\zeta$ is an increasing function of the temperature with a tendency to saturate for large $T$.

For values $0 < d < d_c$
the curves bend downwards, signaling a slower logarithmic growth. 

In the upper set of Fig. (\ref{m2_eps1}) some of the curves displayed in the main picture
are plotted against $\ln t$ on a double logarithmic scale. In this plot a logarithmic 
law $L(t)\sim [\ln t]^\psi$ looks like a straight line with slope $1/\psi$. 
Here one sees that, while the curves for $d=0$ and $d=d_c$ tend to bend upward,
signaling a growth faster than a logarithmic one, the ones for $d=0.1$ and $d=0.25$
can be interpreted as slowly converging to a straight line. 
For completeness we mention that, fitting the data in the last
decade with the above logarithmic form, we find $1/\psi=2.5,2.4,2.1,2.5$ for 
$d=0.05, 0.1, 0.25, 0.4$ respectively. These values, however, should be taken
as a qualitative indication,
due to the difficulty to fit logarithmic forms.

In order to better illustrate this pattern of behaviors, we have computed 
the {\it effective exponent} $1/\zeta_{eff}$ which is obtained by fitting
the curves to a power-law in the last time decade. This is
a tool to compare the speed of growth of the different curves since, 
as we noticed already, a power-law is only observed at $d=0$ and $d=d^*$.
The effective exponent is plotted in the inset of the figure showing
very clearly the non-monotonous behavior of the growth-law as $d$ is varied.

Finally, it should be mentioned that at $d=d_c$ one has $T_c=0$, 
hence our quenches are
in this case above the critical temperature. 
Therefore we expect that the system will eventually reach the final equilibrium
state in a finite time even for ${\cal L}\to \infty$ and the coarsening phenomenon 
we are observing is not a truly asymptotic behavior. 
However, as discussed in 
\cite{diluted}, this preasymptotic stage lasts for a time that diverges in the
$T\to 0$ limit, similarly to what happens in the one-dimensional Ising
model \cite{Claudio,Ontheconnec}. 

\subsection{$\epsilon <J_0$} \label{epslt1m2}

\begin{figure}[h]
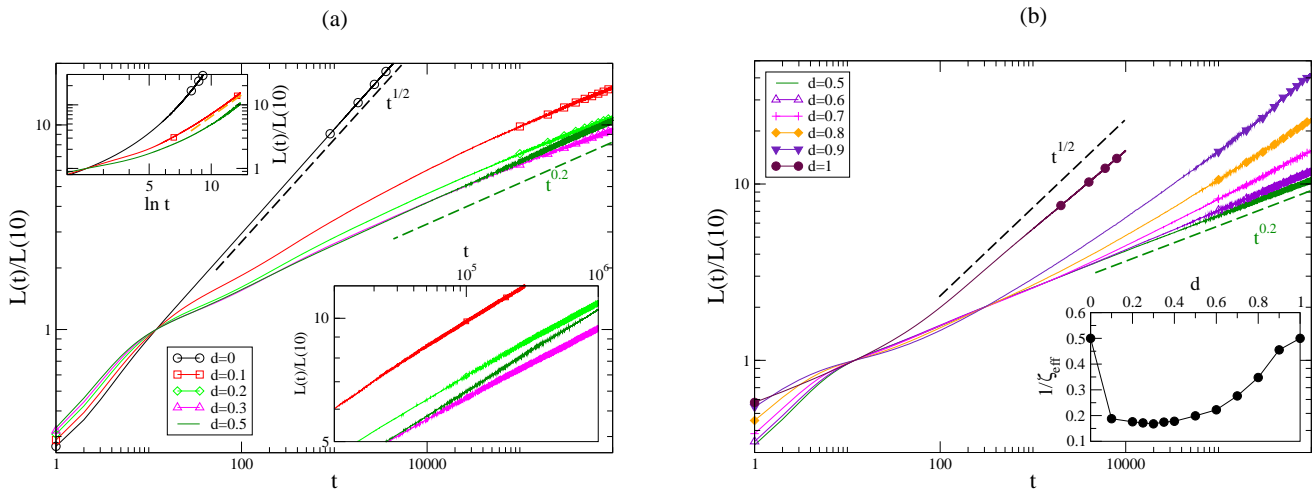

  \vspace{1cm}
    \centering
    %\vbox to 8.5 cm {                                                          
\rotatebox{0}{\resizebox{.45\textwidth}{!}{\includegraphics{m2_eps075_a.eps}}}
\hspace{1cm}
\rotatebox{0}{\resizebox{.45\textwidth}{!}{\includegraphics{m2_eps075_b.eps}}}
\caption{(Color online). $L(t)$ is plotted against $t$ in a double-log plot 
for a quench of the ARBIM to $T=\epsilon$, with $\epsilon =3/4 \,J_0$ 
and different values of $d$ specified in the key. Panel (a) (left) shows the
cases with $0\le d\le d_c$, the right one those with $d_c\le d\le 1$.
The inset is a zoom on the curves in the last two decades. 
The black-dashed line is the
pure-case power-law $L(t)\propto t^{1/2}$. The green-dashed line
is the power-law $L(t)\propto t^{0.2}$.
In the upper inset some of the curves of the main picture are plotted against
$\ln t$ on a double-log plot. The orange-dashed line is the fit  
$y=0.03 \cdot [\ln t]^{2.4}$.
Panel (b) (right) shows the cases with $d_c\le d\le 1$. 
In the inset the behavior of the effective exponent $1/\zeta _{eff}$
is shown for the whole range of $0\le d\le 1$.
%Fits: Red: 0.027247*x^2.4078; Green: 0.029125*x^2.2554; Magenta: 0.033253*x^2.1499;
%DarkGreen: 0.012981*x^2.5485
}
\label{m2_eps075}
\end{figure}

Evidences supporting an interpretation in terms of  the existence of three fixed points along the 
$\epsilon =J_0$ axis were discussed in the previous Section.
Here we show that a similar structure is observed for any value of $\epsilon$. 
In order to do that we have scanned the parameter space by a set
of cuts along lines with fixed $\epsilon$. An example (with $\epsilon =3/4 \,J_0$) 
is shown in Fig. \ref{phas_diag} with an horizontal dashed magenta line.
The behavior of $L(t)$ along this line is reported in Fig. \ref{m2_eps075}.
Even if, when $\epsilon <J_0$,
asymptotic coarsening occurs for any value of $d$ (the brown region
in Fig. \ref{phas_diag} is peculiar to the case with $\epsilon =J_0$) for sufficiently
low temperatures,
in order to compare the data to those for $\epsilon =J_0$ discussed in the
previous Section \ref{eps1m2} (Fig. \ref{m2_eps1}) let us initially restrict the
discussion to the cases with $d\le d_c=0.5$, which are shown in the left panel of
the figure.
For these values of $d$  we see a pattern qualitatively similar to that observed for 
$\epsilon =J_0$ 
(Fig. \ref{m2_eps1}): 
Starting from the power-law $L(t)\sim t^{1/2}$ of the pure case $d=0$,
the growth of $L(t)$ slows down up to a certain value $d^*(\epsilon)\simeq 0.3$ 
and
then speeds up again reaching at $d=d_c=1/2$ a form compatible with a power-law $L(t)\sim t^{1/\zeta (\epsilon, T,d)}$ with an exponent
$1/\zeta (\epsilon=3/4\,J_0/2,T=\epsilon ,d)\simeq 0.2$ roughly similar to the one
$1/\zeta (\epsilon=J_0,T=\epsilon,d_c)\simeq0.22$ observed for $\epsilon=J_0$.
The most notable difference with the case
$\epsilon =J_0$ is the somewhat larger value of $d^*$, since we have 
here $d^*(\epsilon =3/4 \,J_0)\simeq 0.3 > d^*(\epsilon=J_0)\simeq 0.25$. 
The fact that $d^*$ is closer to $d_c$ than for $\epsilon =J_0$ makes the 
corresponding curves closer and the non-monotonic behavior less 
evident than in Fig. \ref{m2_eps1} (this can be better appreciated in the lower inset
of the left part of Fig. \ref{m2_eps075}, 
where a magnification the large-time portion of the
figure is shown, or by inspection of the effective exponent $1/\zeta _{eff}$
reported in the inset of the right panel of the figure). 
In the upper set of Fig. (\ref{m2_eps1}) some of the curves displayed in the main picture
are plotted against $\ln t$ on a double logarithmic scale since, as explained in Sec. \ref{eps1m2},
this is a useful representation to check for a logarithmic growth-law.
While the curves for $d=0$ and $d=d_c$ bend upward,
signaling a growth faster than a logarithmic one, the one for $d=0.1$
can be interpreted, also in this case, as converging to a straight line. 
Fitting all the data in the last
decade with the logarithmic form of Sec. \ref{eps1m2}, one finds 
$1/\psi=2.4,2.3,2.1$ for $d=0.1, 0.2, 0.3$ respectively.

Clearly, although the overall behavior is similar, the interpretation of the cases 
with $\epsilon <J_0$ cannot follow literally the one for $\epsilon =J_0$, which was relying
on the geometrical properties of a diluted system. However, this very same pattern of behaviors
suggests that, since at low temperatures interfaces are located preferentially on the
{\it weak} bonds $J_{ij}=J_0-\epsilon$, what really matters is the topology of the network
of this set of bonds. At $d=d_c=1/2$ such network is a percolation fractal and this turns the
logarithmic growth into a power-law one.

Let us now go back to the data with $d>d_c=1/2$ (still with $\epsilon =3/4J_0$) 
where, at variance with the case 
$\epsilon =J_0$ discussed in Sec. \ref{eps1m2}, 
asymptotic phase-ordering may still occur.
In this region one observes in the right panel of Fig. \ref{m2_eps075}
that the curves bend upward as time elapses. Furthermore 
the growth-law becomes faster and faster as
$d$ is raised from $d=d_c$ up to $d=1$. 
This can be interpreted as follows:
Right at $d=1$ the system is 
a pure one with $J_{ij}\equiv J_0+\epsilon$ and the power-law $L(t)\sim t^{1/2}$
holds. This is analogous to the case with $d=0$, apart from a trivial shift of
the value of the coupling constants. However, starting from this pure case with $d=1$, 
the effect of lowering $d$ is very different from the one occurring
when, starting from the pure case $d=0$, $d$ is increased.
In the former case by lowering $d$ one introduces
{\it strong} bonds $J_{ij}=J_0+\epsilon$ in a majority background of week ones $J_{ij}=J_0-\epsilon$. 
Instead, in the second case, by increasing $d$ one does precisely the opposite, 
introducing strong bonds in a background of weak ones.
Although these two situations might look symmetrical on purely geometrical grounds,
they are not so from the energetic point of view. In fact a weak bond acts as an
attractive pinning point for a wandering interface and an activation energy is 
required to leave it. This brings in the slow logarithmic growth-law for $0<d<d_c$.
Conversely, for $d>d_c$ the strong 
bonds form finite repulsive regions which the interfaces manage to
overcome without activation. Therefore, in the whole region with
$d>d_c$ coarsening is expected to proceed over the compact network of weak bonds
with the pure asymptotic growth-law $L(t)\propto t^{1/2}$. 

Clearly, for $d$ larger than $d_c$ but sufficiently close to $d_c$,
we expect a crossover phenomenon with a growth-law initially influenced 
by the presence of the nearby $W$ line.
The pure-like growth-law should set in when the domains size $L(t)$
has grown larger than the typical size of the clusters of strong bonds  
$J_{ij}=J_0-\epsilon$. Hence the crossover to the late $L(t)\propto t^{1/2}$
behavior should occur earlier upon increasing $d$.

These expected behaviors are consistent with what we observe in
Fig. \ref{m2_eps075}. The data show a gradual crossover as $d$ is increased
above $d_c$ which occurs earlier raising $d$.
The upward bending of the curves shows that the effective exponent is 
bound to increase above the values measured in the time accessed in the simulations.
For $d=0.9$ 
the growth-law is asymptotically very
close to the pure one, as indicated by the effective exponent $1/\zeta _{eff}\lesssim 0.46$
(see inset) which grows up to $1/\zeta _{eff}\simeq 0.47$
if computed for $t>5\cdot 10^5$.

Repeating the simulations for different
values of $\epsilon $, in the whole range $0<\epsilon\le J_0$, we find a 
structure similar to the case $\epsilon =3/4 J_0$ discussed insofar, with a value of $d^*$ which steadily increases as $\epsilon $ is lowered
(specifically we find $d^*(\epsilon=J_0/2)\simeq 0.35$, $d^*(\epsilon=J_0/4)\simeq 0.4-0.45$)). 
This can be interpreted as a line of attractive
fixed points, pictorially drawn 
dot-dashed in red in Fig. \ref{phas_diag},
which, starting from $S$ reaches the $d$ axis. Next to it, there is also
a line of W-like points which starts from $W$ (depicted in green). Along
this line the two kinds of bonds ($J_{ij}=J_0\pm \epsilon$) are arranged
on a percolation network and $L(t)\sim t^{1/\zeta(\epsilon,T,d)}$, similarly
to what occurs at W. 
Concerning the exponent $\zeta (\epsilon,T,d)$ we have found
$1/\zeta \simeq 0.2$, roughly irrespectively of the value of $\epsilon$,
suggesting that this exponent depends sensibly only on the ratio $\epsilon / T$.
The power-law behavior observed at $d_c$ and the properties of the exponent 
$\zeta $ can be understood by considering 
what is observed on a simple model based on a deterministic fractal network, 
as we will discuss in Sec. \ref{m2gas}. 

\subsection{Summary of the results for the ARBIM} \label{firstsum}

Let us briefly summarize
the results for the ARBIM. First of all our results confirm the existence 
of a rich interplay between the behavior of the pure system and 
two other types of dynamical behaviors caused by the inhomogeneities, 
which are characterized
respectively by logarithmic or by temperature-dependent 
power-law growth of $L(t)$.
This can be interpreted along the lines of what has been previously found 
for the SDIM:
Moving along the segment with $\epsilon =J_0$ 
(turquoise in Fig. \ref{phas_diag}) one encounters three fixed points, 
the pure one at $d=0$ and those denoted with
the letters $S$ and $W$.
$W$ is located at the percolation point $d_c=0.5$ and
$S$ at a value $d^*(\epsilon=J_0)\simeq 0.25$ roughly half-way between $d=0$
and $d=d_c$. At $W$ the topology of the substrate is such that the critical
temperature of the Ising model defined on it is $T_c=0$ and this, according to
the conjecture proposed in \cite{lastparma} implies a power-law growth 
$L(t)\sim t^{1/\zeta}$. Since $W$ is a repulsive fixed point the dynamics 
is always ruled by the attractive point $S$ for any $0<d<d_c$,
resulting in an asymptotic logarithmic growth. However the power-laws associated
to the pure fixed point with $d=0$ or to $W$ may show up preasymptotically
due to a crossover phenomenon.

We have found that this same pattern persists also for $\epsilon < J_0$.
Hence, this means that $S$ 
and $W$ are not isolated fixed points, but that each of them belongs to 
two lines of fixed points extending from $\epsilon =J_0$ down to $\epsilon =0$.
In particular, the $W$ line appears to be located at $d=0.5$, irrespective
of $\epsilon $, and is characterized by the asymptotic 
power-law $L(t)\sim t^{1/\zeta}$. This means that in the low temperature
limit what matters for
power-law growth is that the network of the two kind of bonds 
are at the percolation point. An argument developed for a deterministic 
fractal structure supporting the existence of a power-law growth 
also for $\epsilon \neq J_0$ will be presented in 
Sec.  \ref{GAS}. The location of $d^*$, on the other hand, is found to depend
on $\epsilon$ and to approach $d_c$ as $\epsilon \to 0$, as pictorially shown
in Fig. \ref{phas_diag}. 

In agreement with the above structure,
the growth-law has been found to display the following features:

\begin{itemize}

\item{i) Quenches to a region where the system is pure (blue lines in Fig. \ref{phas_diag}): Pure power-law behavior $L(t)\propto t^{1/2}$ from an early 
microscopic time $t_{early}(\epsilon,T,d)$ onwards.}

\item{ii) Quenches to the S-line (dot-dashed red in Fig. \ref{phas_diag}):
Logarithmic behavior $L(t)\propto (\ln t)^{1/\psi}$ from the 
microscopic time $t_{early}(\epsilon,T,d)$ onwards.}

\item{iii) Quenches to the W-line (continuous green in Fig. \ref{phas_diag}):
Disordered power-law behavior $L(t)\propto t^{1/\zeta(\epsilon,T,d)}$ from 
$t_{early}(\epsilon,T,d)$ onwards. Our data suggest that $\zeta $ depends sensibly only
on the ratio $\epsilon /T$.}

\item{iv) Quenches to a point with disorder on the left of the S-line:
A crossover from an early pure behavior $L(t)\propto t^{1/2}$
for $t_{early}(\epsilon,T,d)<t<t_{cross}(\epsilon,T,d)$ to logarithmic behavior
for $t>t_{cross}(\epsilon,T,d)$ (with $t_{cross}(\epsilon,T,d)$ decreasing towards 
$t_{early}(\epsilon,T,d^*(\epsilon))$ 
upon approaching the S-line).} 

\item{v) Quenches between the S and the W-line:
A crossover from an early disordered power-law behavior 
$L(t)\propto t^{1/\zeta (\epsilon,T,d)}$
for $t_{early}(\epsilon,T,d)<t<t_{cross}(\epsilon,T,d)$ 
to logarithmic behavior
for $t>t_{cross}(\epsilon,T,d)$ (with $t_{cross}$ increasing upon approaching
the $W$-line).} 

\item{vi) Quenches to a point with disorder on the right of the W-line
(only for $\epsilon <J_0$):
A crossover from an early disordered power-law behavior $L(t)\propto t^{1/\zeta(\epsilon,T,d)}$ for $t_{early}<t<t_{cross}(\epsilon,T,d)$ to the pure behavior
$L(t)\propto t^{1/2}$
for $t>t_{cross}(\epsilon,T,d)$ (with $t_{cross}(\epsilon,T,d)$ decreasing 
towards $t_{cross}(\epsilon,T,d=1)$ moving away from the
$W$-line).} 
 
\end{itemize}

We mention that the pattern of asymptotic behaviors summarized above is 
a low-temperature feature.

\section{Deterministic fractal networks} \label{GAS}

In this section we study two deterministic fractal networks which can be considered
as a simple paradigm explaining the ARBIM dynamical behavior. 
The structures that we will consider
are generalizations of the Sierpinski gasket (SG) and of the Sierpinski 
carpet (SC). As we will discuss below, these networks can be considered 
representative of structures with a vanishing or with a finite critical temperature,
respectively.

\begin{figure}[h]
  \vspace{2.5cm}
    \centering
    %\vbox to 8.5 cm {                                                       
\vspace{0cm}
\rotatebox{0}{\resizebox{.35\textwidth}{!}{\includegraphics{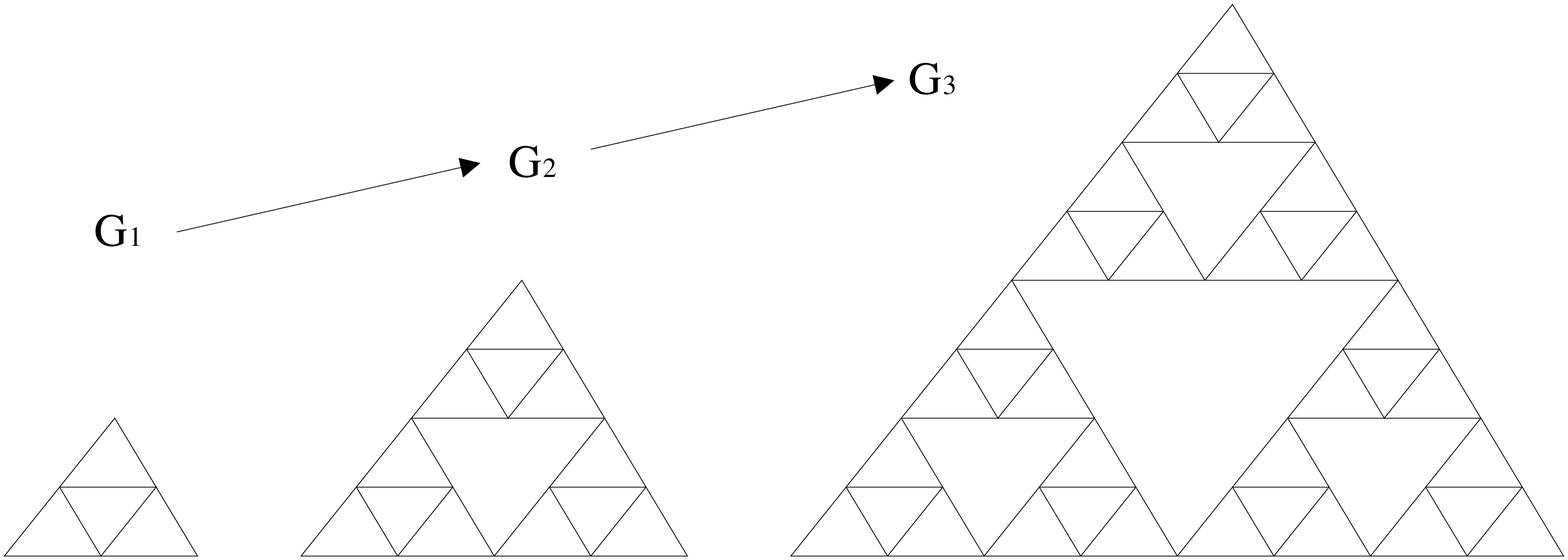}}}
   
\vspace{2.5cm}

\rotatebox{0}{\resizebox{.35\textwidth}{!}{\includegraphics{carpet.eps}}}   

\vspace{2.5cm}

\rotatebox{0}{\resizebox{.35\textwidth}{!}{\includegraphics{sierpinsky.eps}}}

\caption{(Color online). Upper panels: Recursive construction of the SG and of the SC.
Lower panel: The SG (in black) embedded in a triangular lattice
(dashed-red). Bonds marked with a blue X are the cutting bonds.}

\label{sierpinsky}
\end{figure}

The SG is a network that 
can be built recursively as shown in Fig. \ref{sierpinsky} (upper panel).
Starting from a primitive three-spin triangular object, the so called generation-one,
the second generation is built by merging three generation-one structures,
and the procedure is then repeated many times iteratively.
As it can be seen in the lower panel of the figure the SG    
can be embedded on a triangular lattice. In this picture the bonds of the
lattice belonging to the fractal network are plotted in black while the remaining ones,
in the voids of the SG, are red. In this way the SG can be regarded as 
a triangular lattice with a deterministic bond dilution. Notice that the 
SG is a {\it weak} structure, in the sense previously discussed for the 
percolation network, because the removal of a finite number of
links can serve to disconnect arbitrarily large parts of the structure
(an example of such {\it cutting bonds} is shown in Fig. \ref{sierpinsky}).
For this reason the critical temperature of the Ising model on the SG is
$T_c=0$. 

The SC \cite{Gefen} is a structure similar to the gasket, for which
the recursive construction starts with a primitive four-spin square object,
as shown in Fig. \ref{sierpinsky}.
Similarly to the SG it can be embedded in a two-dimensional regular lattice,
although squared instead of triangular.
The most important difference between the SC and the SG is that cutting bonds,
which are present in the former, are absent in the latter.
Related to that, the critical temperature of the Ising model defined
on the SC is finite.

We now introduce a spin model on the embedding lattices of these
structures (either triangular or squared) 
by defining Ising variables
governed by the Hamiltonian (\ref{ham}).
The coupling constants $J_{ij}$ take the value $J_0+\epsilon$ on the 
primary network (either the SG or the SC), 
and $J_0-\epsilon $ in the {\it voids} (namely the dashed-red bonds in Fig. \ref{sierpinsky}) and similarly for the SC.
Clearly, for $\epsilon\equiv0$ one recovers the usual Ising model on the 
homogeneous (triangular or squared) lattice and $\epsilon =J_0$ corresponds to the
Ising model defined on the original SG and SC fractal networks.

The dynamics is that described in Sec. \ref{timev}.
The models introduced above will be denoted in the following as the generalized 
Sierpinski gasket (or carpet) Ising model (GSGIM or GSCIM). 

\subsection{Simulation details}

We have performed a series of numerical simulations of the deterministic
fractal models described above by considering quenches to different 
final temperatures for a SG of $N= 2100225$ sites, corresponding to a fractal 
network of the 11-th generation, and on a SC of $N=2125764$ sites, corresponding
to a fractal of 7-th generation.
Data are averaged over 2-500 independent runs. The other details of the
simulations are as described in Sec. \ref{SIMUL} for the ARBIM.

\subsection{$\epsilon \equiv J_0$}\label{oldsg}

Let us start by recalling what is known for the phase-ordering of the
Ising model with a coupling constant $J$ on the usual SG. Since this is a diluted structure
with $T_c=0$ it can be considered to be comparable in some sense
to the GSGIM with the choice $J_0=J/2, \epsilon =J_0$, namely at the point W.
The Ising model on the SG was studied in Ref. \cite{parma,lastparma} 
where it was shown that $L(t)$ grows
as a power-law with a temperature dependent exponent 
$L(t)\simeq t^{1/\zeta(T)}$. This has to be compared to what observed
in the ARBIM at the fixed point W where, as shown in Sec. \ref{eps1m2},
a similar power-law is observed.
For low temperatures the algebraic behavior is decorated by a periodic oscillation
which, as discussed in \cite{parma,lastparma}, is related to the recursive construction
of the network. We notice, by the way, that similar oscillations are observed
also in disordered models, such as the ARBIM considered above,
at sufficiently low temperatures (lower than the ones used for simulations
in this paper). In \cite{lastparma} it was shown that the algebraic behavior can be
understood in terms of the scaling of the barriers which tend to pin
the displacement of interfaces.
In the following
we briefly sketch the argument (more details can be found in 
Ref. \cite{lastparma}).

\begin{figure}[h]
  \vspace{2.5cm}
    \centering
    %\vbox to 8.5 cm {                                                       
\vspace{0cm}
\rotatebox{0}{\resizebox{.45\textwidth}{!}{\includegraphics{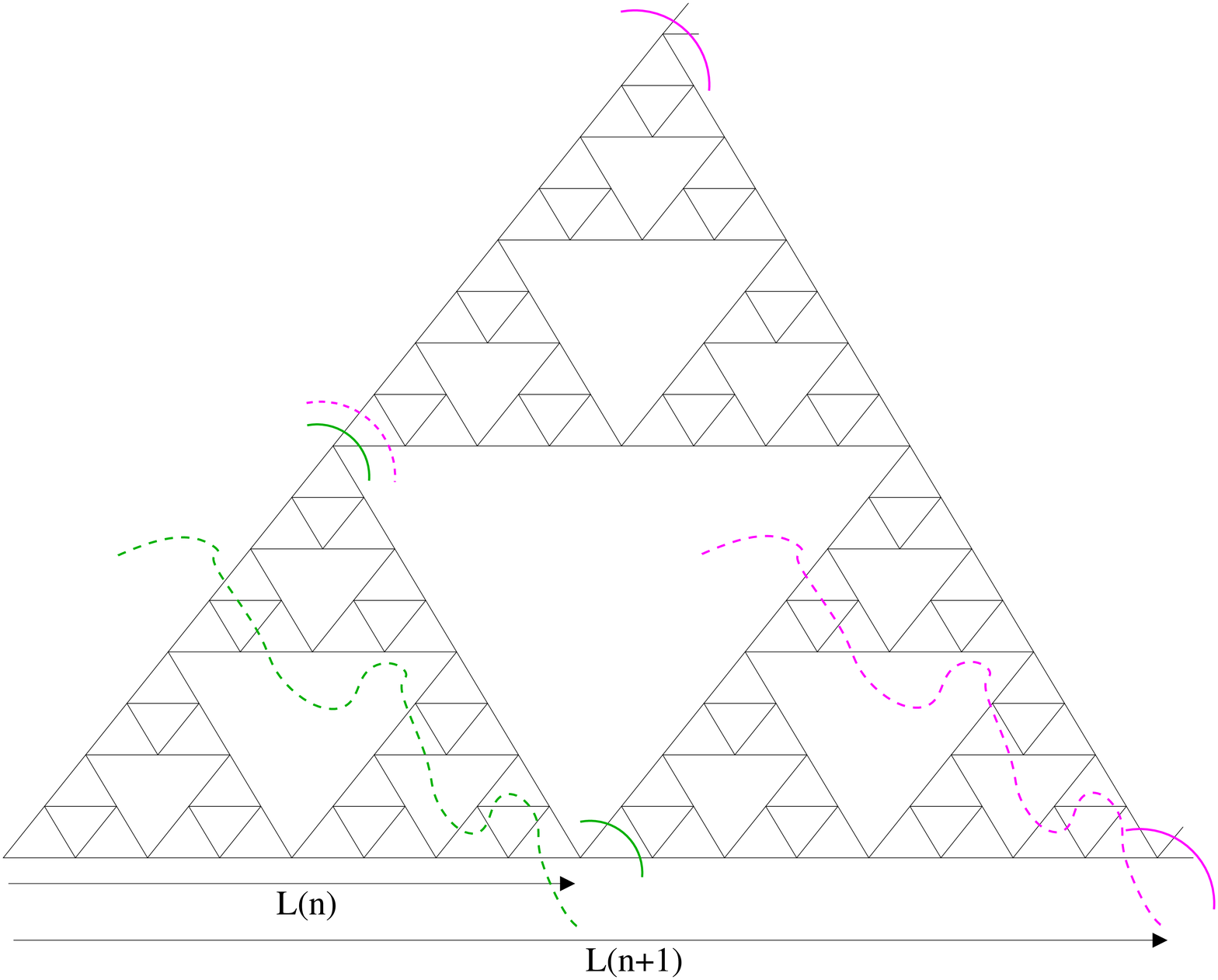}}}
\caption{(Color online). Sketch of the movement of an interface on the SG.}

\label{sierpinski2}
\end{figure}

Fig.~\ref{sierpinski2}  illustrates pictorially the evolution of an interface
which progressively spans a part of the structure. Initially, the position of the 
domain wall 
is outside the figure, in the left corner. This means that
all the spins are, say, down.
As time goes on, the interface enters the graph by moving across the intermediate position
$I_n^{(max)}$ indicated by a dotted green line 
(this means that spins on the left of the green line 
have been reversed up). The index $n$ refers to the fact that the interface is currently spanning the $n$-th generation of the fractal network.
When the spins of the whole generation $n$ have been reversed, the interface,
depicted with two green continuous arches in the middle of the sructure, 
is located in the configuration 
$I_n^{(min)}$ on the four {\it cutting bonds}. 
Since the energy of a domain wall is $2J$ times its length, it is clear that
in $I_n^{(min)}$ the interface has a minimum energy $E_n^{(min)}=8J$
(notice that this quantity is independent of $n$). 
Let us assume that the highest energy $E_n^{(max)}$ of the system during the above process was reached
in the (generic) configuration $I_n^{(max)}$, so that a barrier of height 
${\cal E}_n=E_n^{(max)}-E_n^{(min)}$ has been crossed.
Now the interface must proceed again to the right in order to reverse all the spins
of the next generation $n+1$, thus reaching the position $I_{n+1}^{(min)}$ located on the other four {\it cutting bonds},
and indicated by the couple of continuous magenta arches on the upper and
lower right corners of the structure 
(all the spins in the figure at this stage have then been reversed). 
Also this configuration has an energy equal to $E^{(min)}$.
This configuration can be reached by sequentially reversing parts of
generation $n$ of the structure (e.g. by first reversing another triangle of
generation $n$, say the lower-right one in Fig. \ref{sierpinski2}).
This event is analogous to the one described before.
In particular, the interface in the intermediate position $I_{n+1}^{(max)}$, 
depicted with a dashed-magenta line on the right of the structure, 
is analogous to the previous one at $I_n^{(max)}$
(dotted-green), except for the presence of an extra part which in the present 
example is indicated with a dashed-magenta arch in the middle of the left side.
Denoting with $E_{n+1}^{(max)}$ the maximum energy reached by the
system in the reversal of the $n+1$ generation, 
one concludes that $E_{n+1}^{(max)}=E_n^{(max)}+4J$, where $4J$ is the 
extra amount of energy
due to the new part of interface (the dotted-magenta arch
in the middle of the left side in the figure). Then
${\cal E}_{n+1}\simeq {\cal E}_n+4J$.
Rewriting ${\cal E}$ in terms of the size $L_n\simeq 2^n-1$ of the 
$n$-th generation, using $L_{n+1}\simeq 2\cdot L_n$, one has
${\cal E}(2\cdot L_n)\simeq {\cal E}(L_n)+4J$.
From this relation, dropping the index $n$, one has
${\cal E}\simeq \frac{4J}{\ln 2}\ln L$.
Assuming an Arrhenius time 
$t\propto e^{\frac{{\cal E}}{K_BT}}$ to overcome the barrier, we arrive at a algebraic 
growth law 
$L(t)\sim t^{\frac{1}{\zeta (T)}}$
with
\be 
\zeta \simeq \frac{4}{k_B \ln 2}\,\frac{J}{T},
\label{stimazeta}
\ee 
at low temperatures. The argument has been developed for the particular
SG structure but it is expected to hold more generally
for any finitely ramified deterministic fractal network and 
also for disordered structures with $T_c=0$, 
since the key ingredient is their {\it weakness},
namely the possibility to disconnect an arbitrary large part of the network 
by cutting a finite number of links. Due to that, the energy barriers
that trap the interfaces on the cutting bonds
grow as slow as a logarithm of the size of the domains.

A completely different situation is encountered when the Ising model is
defined on the SC. In this case, due to the absence of cutting bonds,
an argument similar to one presented above shows that the pinning 
energy of the interfaces grows faster - in an algebraic way - with the
size of the ordered domains. Details can be found in~\cite{lastparma}. 

\subsection{$0<\epsilon < J_0$} \label{m2gas}

\begin{figure}[h]
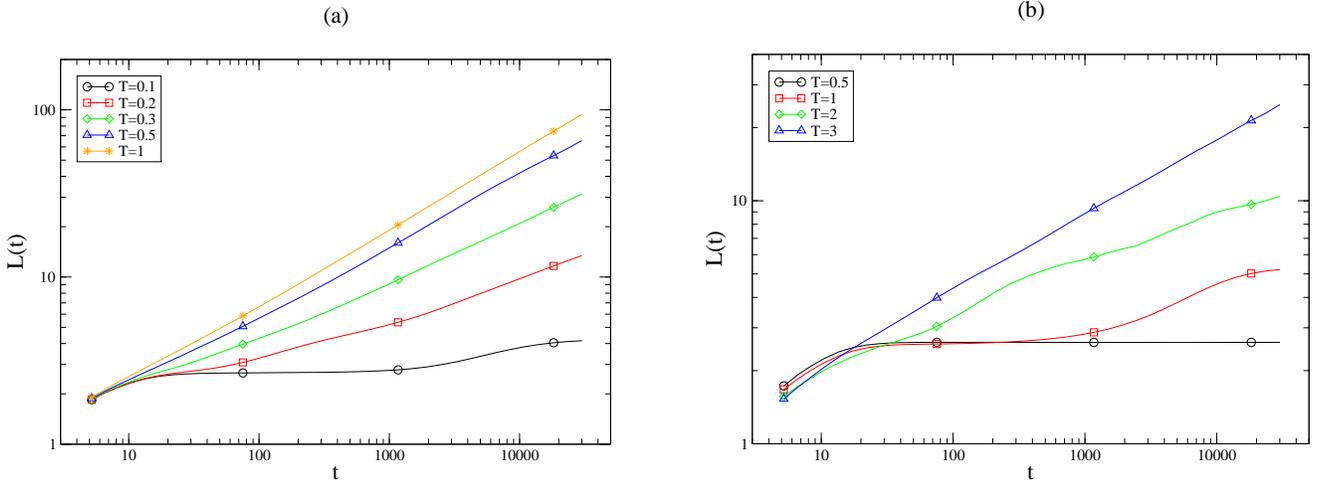

  \vspace{2.5cm}
    \centering
    %\vbox to 8.5 cm {                                                       
\vspace{0cm}
\rotatebox{0}{\resizebox{.45\textwidth}{!}{\includegraphics{epsilon_01.eps}}}
\hspace{1cm}
\rotatebox{0}{\resizebox{.45\textwidth}{!}{\includegraphics{epsilon_09.eps}}}
\caption{(Color online). $L(t)$ for the GSGIM with $\epsilon/J_0=10^{-1}$ (panel (a), left) and 
$\epsilon/J_0=0.9$ (panel (b), right).}
\label{gasm2}
\end{figure}

We have shown in Sec. \ref{epslt1m2} that in the ARBIM the characteristic power-law behavior 
$L(t)\sim t^{1/\zeta}$ of the point W
is observed not only at W but along the whole line with $d=1/2$, 
the only effect of changing $\epsilon$ is possibly to change the exponent $1/\zeta$. 
Our analogy suggests that the same behavior should be observed also in the GSGIM.
We have performed a set of simulations with a small value of 
$J_0/\epsilon$ ($\epsilon /J_0=10^{-1}$)
and another for a much larger value of this parameter ($\epsilon /J_0=0.9$). For each
choice of $\epsilon$ we have considered different values
of the quenching temperature $T$.
The results of our simulations are shown in Fig. \ref{gasm2}.

Starting with the case with $\epsilon/J_0=10^{-1}$ (left panel of Fig. \ref{gasm2}) 
we see that, as expected, the growth of $L(t)$ is 
similar to the one observed in the Ising model on the SG (namely the case with 
$\epsilon/J_0=1$), namely a power-law decorated by periodic oscillations
whose amplitude increases upon lowering $T/\epsilon$.
This behavior is also analogous to what observed in the ARBIM for $d=1/2$
(except for the oscillations), where the
network of bonds with $J_{ij}=J_0+\epsilon$ is a percolation fractal for which
the Ising critical temperature vanishes as in the SG. 

For $\epsilon/J_0=0.9$ the situation is qualitatively similar, with the quantitative
difference of a much slower growth, for a given temperature, with respect to the
case with for $\epsilon/J_0=0.1$. This is expected since the height of barriers is
related to $\epsilon $. However, comparing curves
for $\epsilon/J_0=0.9$ and $\epsilon/J_0=10^{-1}$ with a similar value of 
$\epsilon /T$, one finds roughly similar exponents (for instance
for $\epsilon/J_0=0.1, T=0.2$ and $\epsilon/J_0=0.9, T=2$ 
one has $1/\zeta \simeq 0.23$ and $1/\zeta \simeq 0.21$ respectively,
and for $\epsilon/J_0=0.1, T=0.3$ and $\epsilon/J_0=0.9, T=3$ 
one has $1/\zeta \simeq 0.34$ and $1/\zeta \simeq 0.31$ respectively).      
This shows that the growth-law depends at most weakly on the parameter $\epsilon /T$.

In order to understand the origin of these behaviors we go back to the argument
of Sec. \ref{oldsg} for the Ising model on the SG.  
When the interface is in the configuration 
$I_n^{(min)}$ sketched with two continuous-green arches in the middle of the structure in 
Fig. \ref{sierpinski2}, the lower-left triangle ABC of 
Fig. \ref{sierpinsky} contains up spins and the rest of the structure is down
(and there are no spins in the voids). The next spin to be reversed is one of
those marked with a bold-green circle. Since these are equivalent, to be 
concrete let us chose it to be the one denoted by the letter p.
The energy needed to flip it is easily computed to be 
\be
{\cal E}_p=4J
\label{enpsg}
\ee
In the case of the GSGIM the
dynamics proceeds along the same lines. However, starting again with 
the interface in the configuration with two continuous-green arches, 
besides the spin marked with
the green circle in Fig.~\ref{sierpinsky} also those indicated by
a heavy violet box can flip.    
Computing the energetic barriers one can easily 
convince that one of the latter is
the next spin to be reversed (for which the 
energy barrier is lower). Let us stipulate this to be 
the one denoted with the letter q.
The energy cost
turns out to be $4(J_0-\epsilon)$. 
After reversing q, spin p
will be flipped, which requires an energy $8\epsilon$.
The maximum energy required to arrive to flip p is then
\be
{\cal E}_p^{GSGIM}=\max[4(J_0-\epsilon),8\epsilon]
\label{enpgsgim}
\ee
At this point we are left back to the situation reached by the 
Ising model on the SG described before, since the blocked spin
p has been flipped, and we can repeat the argument developed before
to arrive to the conclusion that also in this case barriers grow
logarithmically with the size of the domains, which in turn
implies a temperature-dependent power-law growth of $L(t)$.
Notice however that in the case of the GSGIM the reversal of
p has been {\it facilitated} by the intermediate flipping
of q. This, as we will show below, lowers the energy 
of the pinning barriers (still maintaining the
logarithmic scaling with $L(t)$).
In order to show this we must compare the energy scale to flip p in
the GSGIM Eq.~(\ref{enpgsgim}) to 
the one~(\ref{enpsg}) of the Ising model on the SG with a coupling constant 
$J=(J_0+\epsilon)$
(namely, the one obtained from the GSGIM by removing altogether the bonds in
the voids of the SG). The latter turns out to be always higher for 
$\epsilon <J_0$.
Then, we
can conclude that the evolution is faster in the GSGIM or, in other
words, that the addition of the extra dashed-red bonds of Fig. \ref{sierpinsky} 
speeds up the dynamics.
Indeed, for instance, extracting the effective exponent for the case with $\epsilon /J_0=10^{-1}, T=1$ we find $1/\zeta \simeq 0.46$, a much larger value than that found for the 
Ising model on the SG with $J=J_0+\epsilon=1.1$ at the same temperature which is
$1/\zeta \simeq 0.23$.

\begin{figure}[h]
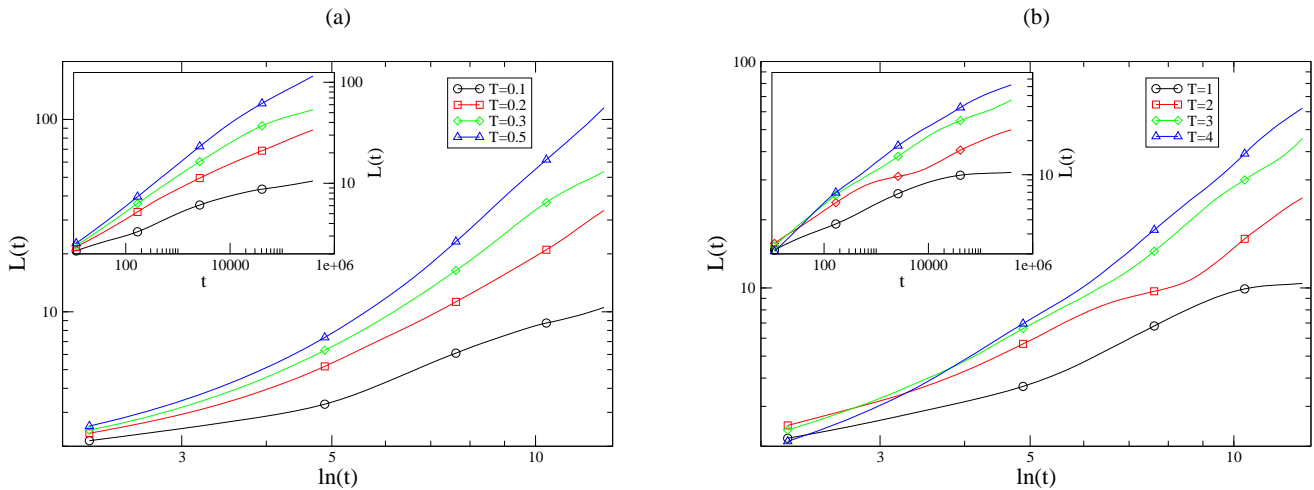

  \vspace{2.5cm}
    \centering
    %\vbox to 8.5 cm {                                                       
\vspace{0cm}
\rotatebox{0}{\resizebox{.45\textwidth}{!}{\includegraphics{epsilon_01_carp.eps}}}
\hspace{1cm}
\rotatebox{0}{\resizebox{.45\textwidth}{!}{\includegraphics{epsilon_09_carp.eps}}}
\caption{(Color online). $L(t)$ for the GSCIM with $\epsilon/J_0=10^{-1}$ (panel (a), left) and 
$\epsilon/J_0=0.9$ (panel (b), right).}
\label{carm2}
\end{figure}

In order check the consistency of our conjectures we have performed
a series of simulation analogous to those for the GSGIM also for the
GSCIM. In this case, since on the SC cutting bonds are absent and
the Ising model defined on it has $T_c>0$, 
we expect to see a situation similar to that 
observed in the ARBIM away from the line of fixed points W,
namely an asymptotic logarithmic law for the domains size.
For the Ising model defined on the SC, in fact, it was shown 
in~\cite{lastparma} that this is indeed the case.
For the GSCIM the results for $L(t)$ are presented in Fig. \ref{carm2}. 
From the inset of this figure one sees that on a double log plot
the data clearly bend downwards for both the values of 
$\epsilon/J_0$ considered. This indicates that the growth of
the domains size is lower than a power-law.
In the main figure the same data for $L(t)$ are plotted 
against $\ln t$, still in a double logarithmic scale.
For a growth like $L\sim (\ln t)^\phi$ in this kind
of plot one should observe a straight line with slope $\phi$.
Due to the oscillating character of the data one cannot
rich a clearcut evidence on the real form of $L(t)$.
However, particularly for the higher values of temperatures
considered, one can conclude that at least a logarithmic
behavior describes much better the data with respect to a power-law.
Notice that, also in this case, curves with comparable values of
$\epsilon /T$ behave quite similarly, suggesting that this combination
of parameters only weekly affects the dynamics.

\subsection{Summary of the results for the GSGIM and analogies with the ARBIM}

In summary, we have shown the following analogies between the ARBIM and the
paradigmatic models based on deterministic fractal networks:

\begin{itemize}
\item{i) For $\epsilon =J_0$, when the randomness of the coupling constant amounts
to bond dilution, one finds a power-law growth of the ordered domains 
(possibly with periodic oscillations) in those cases
with weak diluted networks. This means the ARBIM at the percolation threshold 
or the GSGIM (since the SG has $T_c=0$).}

\item{ii) One finds power-law growth also for $\epsilon <J_0$ in those cases where
the network of the stronger bonds $J_{ij}=J_0+\epsilon $ is weak, i.e. in the ARBIM on the 
green W line schematized in Fig. \ref{phas_diag}, or in the GSGIM.}

\item{iii) The growth-law slower than a power-law (of a logarithmic
type) in all the other cases, namely when the network of bonds 
with $J_{ij}=J_0+\epsilon$ is not weak. This corresponds to the region of 
parameter space of the ARBIM away from the green W-line of Fig. \ref{phas_diag}
or, considering the fractal structures, in the GSCIM for any choice of 
the parameters.}

\end{itemize}

These observations can be rationalized as follows:
The GSGIM depends on the quantities $\epsilon /J_0$, $T/J_0$, 
whereas in the ARBIM
the extra parameter $d$ is also present. The network of bonds $J_{ij}=J_0-\epsilon$
of the GSGIM, being a SG, is always weak. The dynamics of this model can then be
understood considering that, in a sense, changing $\epsilon$ corresponds
to drive the ARBIM along the green W-line. 
On the other hand, the GSCIM, which is based on a structure with no cutting 
bonds, corresponds to the ARBIM away from
the line of fixed points W.

\section{Conclusions and perspectives} \label{CONCL}

This paper was devoted to the study of phase-ordering on inhomogeneous systems,
focusing on the growth-law of the ordered domains. 
Previous studies on systems with site dilution, either random~\cite{diluted} or deterministic~\cite{lastparma},
suggest that the growth-law is strongly affected by the topological
properties of the substrate, namely the network of non-diluted sites, similarly to what 
is known for equilibrium properties. Specifically, it was shown that the growth-law 
is of a logarithmic type if the substrate is such that a finite-temperature phase-transition is 
present, whereas it turns to a to power-law on networks that do not support such phase-transition. 
This suggests that a similar correspondence could be at work also in systems whith
different kinds of inhomogeneities. 

In this paper we have studied this possibility 
by considering, as a first step, a system with bond dilution, a source of inhomogeneity different,
but as close as possible, to that of site-diluted systems. 
Besides addressing the problem of the influence of topology on the non-equilibrium processes,
our study is also ment to possibly shed some light on the nature of the power-law like 
pre-asymptotic regime in ferromagnetic systems with a random bond distribution \cite{CLMPZ,various2}.
In this perspective, we have undertaken a thorough investigation of the non-equilibrium
evolution of a quenched random-bond Ising model, the ARBIM, where two kind of bonds with strength
$J_{ij}=J_0\pm \epsilon$ occurring with relative probabilities $d$, $1-d$, are randomly seeded on
a two-dimensional square-lattice. This allows us not only to study the case of bond dilution,
corresponding to $\epsilon =J_0$, but also the generic case which, as discussed above, is relevant
to the physics of disordered ferromagnets.  

Our study shows that the whole pattern of behaviors of the model as its parameters 
(such as $\epsilon$, the quench temperature, and $d$) are varied can be interpreted in terms either of
logarithmic or power-law growth laws, and of crossover phenomena between them,
similarly to what was found in models with site dilution~\cite{diluted}.
For $\epsilon =J_0$, when the randomness of the bonds amounts to a dilution, this interplay can be 
interpreted in terms of the topology of the substrate. 
Interestingly, this very interpretation provides a framework to 
the behavior of the system also for $\epsilon \ne J_0$, where the form of bond randomness does
not amount to simple dilution.

To check the generality of such an interpretation, and in order to study the phenomenon
on simpler systems, we have also considered a ferromagnetic model defined on deterministic
substrates, such as the Sierpinski carpet and gasket, which do (or do not, respectively) support 
a finite-temperature phase-transition. 
According to the general conjecture put forth in~\cite{lastparma}, these models should represent two
paradigmatic examples of logarithmic and power-law growth, because these behaviors are not 
expected to depend on specific details of the system, but they only rely on the presence of a 
critical point. Since these structures have the advantage
to be deterministic, they can be regarded as toy systems where the properties observed in
the more complex disordered cases, as in the ARBIM model, can be more carefully studied,
checked and, possibly, understood. Indeed, by defining an Ising model with the same bimodal distribution
of bonds as in the ARBIM, we were able to identify a similar pattern of behaviors to the one
observed in the ARBIM.  

Since the occurrence of preasymptotic power-laws followed by a logarithmic growth
is observed also in systems with different kinds
of quenched disorder as, for instance, random fields, it would be
interesting to understand if a similar interpretation applies also in these
cases. This would promote the conjecture put forth in this paper to a more general
character.

Next to the behavior of $L(t)$, the role of
disorder (and more generally of any source of inhomogeneity) is relevant 
for the form of 
the scaling functions of correlation and response functions~\cite{altric}.  
In particular the so-called {\it superuniversality }~\cite{10noirf}, 
namely the
property according to which $L(t)$ should depend on the strength of the
disorder while the scaling functions should not, after some 
initial confirmations~\cite{super1,super2,super3,super} 
has been recently questioned~\cite{diluted,decandia,CLMPZ,variousnoi2}.
In this respect, the results of this paper and their interpretation in 
terms of competition between different scaling behaviors strengthen
previous findings against superuniversality.

\vspace{1cm}
{\bf Acknowledgments}
F.Corberi acknowledges financial support by MURST PRIN 2010HXAW77\_005.
E.Lippiello acknowledges financial support from
the National Science Foundation under Grant No. NSF PHY11-25915.

\end{document}